\documentclass[reprint,amsmath,amssymb,prb,showpacs]{revtex4-1}
\usepackage{rotating}
\usepackage{nicefrac}
\usepackage{multirow}
\usepackage{color}
\usepackage{graphicx}
\usepackage{epsfig}
\usepackage{dcolumn}
\usepackage{bm}
\DeclareGraphicsExtensions{.pdf,.png,.jpg,.eps}

\begin{document}

\title{Guidelines for understanding cubic manganese-rich Heusler compounds}
\author{Lukas~Wollmann,$^{1}$ Stanislav~Chadov,$^{1}$ J\"urgen K\"ubler,$^{2}$ Claudia Felser$^{1}$}
\affiliation{$^1$Max-Planck-Institut f\"ur Chemische Physik fester
        Stoffe,  N\"othnitzer Strasse~40, 01187 Dresden, Germany}
\affiliation{$^2$Institut f\"{u}r Festk\"{o}rperphysik, Technische
        Universit\"{a}t Darmstadt, 64289 Darmstadt, Germany}

\begin{abstract}
Manganese-rich Heusler compounds are attracting much interest in 
the context of spin transfer torque and rare-earth
free hard magnets.  Here we give a comprehensive overview of the magnetic 
properties of non-centrosymmetric cubic Mn$_2$-based Heusler  materials, which
are characterized  by  an   antiparallel  coupling  of magnetic
moments on Mn  atoms.  Such a ferrimagnetic order leads to the emergence
of new properties that are absent in  ferromagnetic centrosymmetric
Heusler  structures. In  terms  of the  band  structure calculations,  we
explain the formation of this magnetic order and the Curie temperatures.
This   overview  is intended   to   establish  guidelines  for   a basic
understanding of magnetism in Mn$_2$-based Heusler compounds.
\end{abstract}

\maketitle

\section{\label{sec:Intro}Introduction}

 Heusler compounds became objects of interest as a class of materials with peculiar transport and magneto-optical properties with 
 the prediction of half-metallicity (gapping in one spin channel) in NiMnSb and Co$_2$MnSn \cite{Groot:1983cg,Kubler:1983fj} in 1983. The whole family of 
 Co$_2${\it YZ} compounds has been considered suitable materials for spintronic devices, and Co$_2$Cr$_{0.6}$Fe$_{0.4}$Al
has from then on been the pioneering candidate material.\cite{Block:2003dt} The potential of this class has been explored in the 
context of magneto-resistance applications, such as in giant-magnetoresistance (GMR) and tunnel-magnetoresistance (TMR) devices. 
Usage as an electrode material for spin injection, where spin polarization is an inevitable prerequisite, and for spintronics applications 
in general\cite{Felser:2013wa} has been realized. Accompanied by the high Curie temperatures shown by many of the compounds in the 
Co$_2${\it YZ} family, the indispensable premises for room-temperature applications are given. On the basis of these premises, TMR ratios 
of 1900\% at low temperatures were accomplished in Co$_2$MnSi$\vert$MgO$\vert$Co$_2$MnSi tunnel junctions by 
Yamamoto {\it et al.}\cite{Liu:2012bp} in 2012, whereas the experimental proof of half-metallicity in Co$_2$MnSi was given by 
Jourdan { \it et al.}\cite{Jourdan:2014hj} just recently. To understand the trends in this class of materials, simple 
rules have been formed from experimental and  theoretical work, and the
Slater--Pauling (SP) behavior of these compounds has been
intensively studied.\cite{Kubler:2007im,Kandpal:2007tn,Galanakis:2002fk}
\begin{figure}
\includegraphics[width=1.0\linewidth]{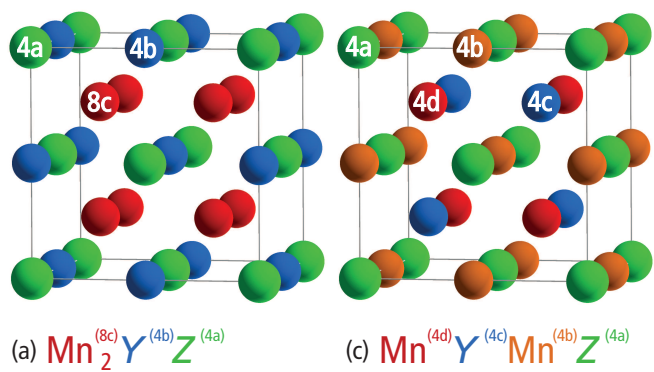}
\caption{\label{fig0} Two basic ordering possibilities for the ternary
        $X_2${\it YZ} composition ($X$, $Y$ - transition metals and $Z$ - main-group element, marked as red, blue, and green, respectively) within the fcc lattice: 
        (a)~L2$_1$-type: occupation of \textsl{8c} by $X$ and  \textsl{4b} by $Y$ (``{\it regular}'' structure);
        (b)~X$_a$-type:  \textsl{8c} is split into \textsl{4c} and \textsl{4d} sites due to their different occupation (``{\it inverse}'' structure).}
\end{figure}

Nowadays, another family of Heusler compounds, the class Mn$_2${\it YZ}, has attracted considerable attention for implementation
 as a free magnetic layer in spin-transfer torque devices such as spin-transfer torque random-access memory (STT-MRAM).\cite{Winterlik:2012cp} In these devices, a spin-polarized current 
 is passed through a hard magnetic layer whose magnetization is switched through transfer of angular momentum.\cite{Slonczewski:1996gq} 
 The most famous member of this group of materials is tetragonal Mn$_3$Ga.\cite{Balke:2007ebba,Wu:2009ep} Starting from its 
 prediction as a compensated cubic ferrimagnet, much research in the field has been invested to promote the implementation of this 
 compound. The reasons are found in its properties, namely, a low experimental magnetic moment, high perpendicular magneto-crystalline 
 anisotropy (PMA) owing to its tetragonal structure, and a high Curie temperature of more than 700 K,\cite{Balke:2007ebba,Winterlik:2008fj} 
which  ensures the thermal stability of the stored information. These properties, in combination with affordable constituent elements, make this 
 material most attractive for high-technology utilization.
 
Despite  difficulties in the realization of such devices, other members of the Mn$_2${\it YZ} family have demonstrated their 
potential.\cite{Ouardi:2013koba,Winterlik:2012cp} Recently, the spin-gapless semiconductor Mn$_2$CoAl was predicted and realized, 
unveiling once again the broad variety of effects to be found in Heusler materials. Peculiar transport properties were expected and have been found 
in such systems, making them promising candidates for room-temperature
semiconductor spintronics.

In addition to the above mentioned materials, the whole family of compounds based on Mn$_2${\it YZ} may show striking properties. 
To optimize these materials, we need a general understanding of how to design compounds with higher spin polarization and achieve compensation of the magnetization in this class. Therefore, we intend to establish guidelines and form simple rules for the 
Mn$_2${\it YZ} family of Heusler compounds.
 
Here we intend to explain the formation of the total moments through monitoring of the local magnetic structure and its influence 
on easily accessible, measurable properties such as the Curie temperature. While the existence of several Heusler-derived structure types 
is well known, \cite{Sticht:1989dc, Kren:1970bm, Niida:1983jb, Kurt:2011hh, Kurt:2012fj, Balke:2007ebba} the interplay between the structure, 
magnetism, and local magnetic moments has not been elucidated as yet, in contrast to the intensively studied cobalt-based Heusler alloys. 
We will present our work based on theoretical methods, i.e.,  density functional theory, focusing on the local  magnetic structure and the 
emergence of the magnetic moments on the Mn atoms as the composition is altered. Furthermore, the evaluation of the interatomic 
exchange in terms of the exchange constant $J_{ij}$ is essential for understanding the formation of the resulting magnetic order. The influence 
of the electron count on the magnitude of the total moments, the composing local moments, and the exchange interaction is disentangled. In a later publication, we intend to apply the gathered knowledge to tetragonal Mn-based compounds.

\section{\label{sec:Cryst} Crystal and magnetic structures}

For ordered cubic Heusler materials, two prototypical structure types
exist. The so-called \textsl{``regular"} type (Cu$_2$MnAl, L2$_1$
prototype) crystallizes in spacegroup (SG) 225, with three inequivalent
Wyckoff positions (\textsl{8c}, \textsl{4b}, and \textsl{4a})
incorporating four atoms per unit cell. These positions are occupied
according to the following scheme. The  \textsl{8c} Wyckoff position
($\frac{1}{4},\frac{1}{4},\frac{1}{4}$) is occupied by two Mn atoms,
whereby  positions \textsl{4b} ($\frac{1}{2},\frac{1}{2},\frac{1}{2}$)
and \textsl{4a} ($0,0,0$) are filled with $Y$ and $Z$ atoms. $Y$
represents any transition metal (TM) and $Z$ stands for a main
group element. The nearest-neighbor coordination can be seen in Fig.~\ref{fig0}\,(a). 
\begin{figure*}[htbp]
\includegraphics[scale=0.32]{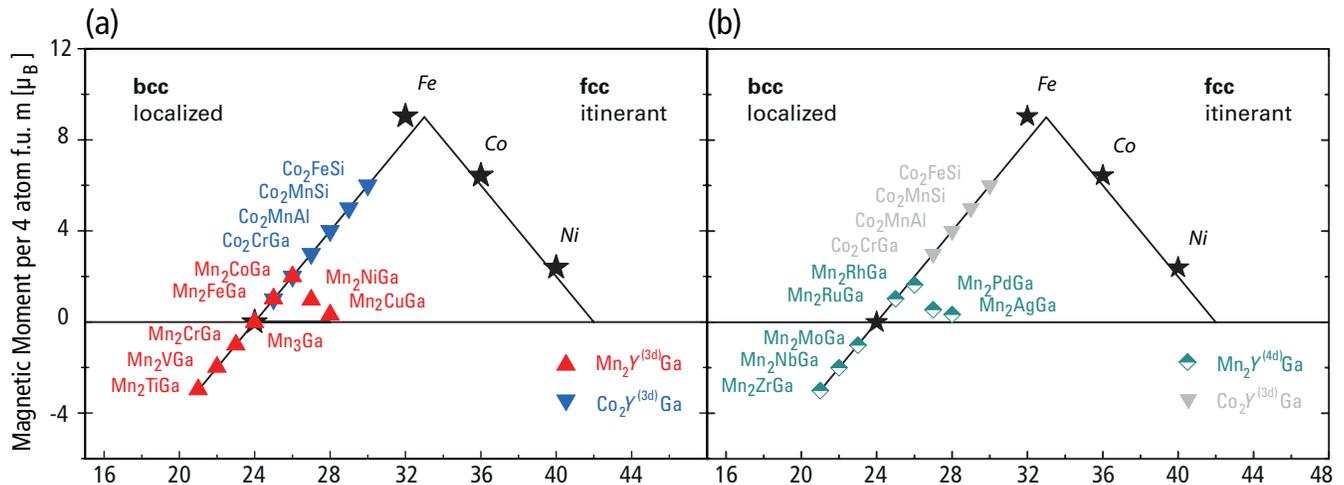}
\caption{\label{fig1-1}SP behavior for examples of (a) Mn$_2$\textsl{Y}$^{(3d)}$Ga and Co$_2$\textsl{Y}$^{(3d)}$Ga compounds and (b) Mn$_2$\textsl{Y}$^{(4d)}$Ga compounds.}.
\end{figure*}
Another prototype often appearing in the context of Heusler compounds
is the Hg$_2$CuTi-type structure (sometimes named \textsl{``inverse"}
Heusler type),  which can be derived from the \textsl{``regular"}
Heusler type by interchanging half of the atoms on position \textsl{8c}
with the \textsl{4b}-position-occupying element. This reduces the
symmetry of the cell, leading to spacegroup 216 (X$_a$-type) with four
inequivalent positions in the unit cell. In this case, these are
occupied with Mn at \textsl{4d} ($\frac{1}{4},\frac{1}{4},\frac{1}{4}$),
$Y$ at \textsl{4c} ($\frac{3}{4},\frac{3}{4},\frac{3}{4}$), Mn at
\textsl{4b} ($\frac{1}{2},\frac{1}{2},\frac{1}{2}$), and $Z$ at
\textsl{4a} ($0,0,0$).

The occurrence of these two types of structures for ternary and even quaternary Heusler alloys has been related to a rule of thumb,\cite{Graf:2011jj} which states for $X_2YZ$ that (i) the inversion-symmetric L2$_1$-type structure with $X$ occupying site \textsl{8c} is favored if \textsl{X} has a higher valence electron count than \textsl{Y}. However, (ii) if \textsl{X} is the earlier transition metal,
then the structure lacks inversion symmetry, as  site \textsl{8c} splits into two inequivalent sites, \textsl{4c} and \textsl{4d}, which are occupied by \textsl{X} and \textsl{Y}, while the second \textsl{X} is found on site \textsl{4b}. Intermixing of sites has been reported, leading to disordered structural variants of the initial Heusler structure types.\cite{Graf:2011jj} Additional structures in the phase space of Heusler compounds are tetragonal derivatives from the cubic parent phases, which can be obtained from these by elongation or compression of the cell axes.

The family of Mn$_2${\it YZ}-based Heusler systems is relatively new as
compared to other groups such as Co$_2${\it YZ}, Fe$_2${\it YZ}, and
Ni$_2${\it YZ}, which have been thoroughly studied over the last few
decades. In the context of magnetism, the distinct feature of Mn$_2$
systems is that their magnetization typically does not exceed
2~$\mu_{\rm B}$/f.u. (in certain rare cases it can reach 4~$\mu_{\rm
  B}$). This is different from the {``\it older}'' groups, where the
magnetization reaches values of 5--6~$\mu_{\rm B}$/f.u. because those
materials are known to exhibit ferromagnetic ordering and to incorporate
elements with higher number of valence $d$ electrons on the $X$ position
(Mn$_3$Ga: $N_{\rm V}=24$, $M=0~\mu_{\rm B}$; Co$_2$MnGa: $N_{\rm
  V}=28$, $M=4~\mu_{\rm B}$), leading to a higher magnetization according to the SP rule. The SP rule for four atoms per formula unit describes the relationship between $N_{\rm V}$ of a compound and its measurable magnetic moment, which is given by
\begin{equation}
M = N_{\rm V} - 24
\end{equation}
This kind of behavior was first  observed for pure \textsl{3d} transition metals and alloys by Slater, Pauling, and Friedel,\cite{Slater:1936fj,Pauling:1938fw,Friedel:1958ab} and it was later  applied to cubic half-metallic Heusler materials.\cite{Kubler:1984bv} The description and origin of the SP rule for several Heusler materials was given earlier.\cite{Galanakis:2002fk,Skaftouros:2013ftba,Graf:2011jj}  The observable types of magnetic ordering in the Mn$_2${\it YZ}-based Heusler family will be part of this work and will be explained in the following sections.

\section{\label{sec:CompDet}Computational Details}
The numerical work was carried out within density functional theory as
implemented in the all-electron FP-LAPW code,
\textsc{WIEN2k}.\cite{wien2k} As exchange- and correlation functional,
the generalized gradient approximation (GGA) in the parametrization of
Perdew, Burke, and Enzerhof was chosen. \cite{Perdew:1996iq} The angular
momentum truncation was set to ${l_{\rm max}=9}$ and ${RK_{\rm
    max}=9}$. The energy convergence criterion was set to $10^{-5}$~Ry,
whereas the charge convergence was set to $10^{-3}$~Ry. All calculations
where carried out on a ${20\times20\times20}$ $k$-mesh, leading to 256
points in the irreducible wedge of the Brillouin zone.
 
For each compound, calculations starting with ferro- and ferrimagnetic configurations were carried out in  space groups No.~225 and 216 according to the regular and  inverse Heusler structure types (Fig.~\ref{fig0}), respectively. The equilibrium lattice constant was determined through volume relaxations of the unit cell. The results of the calculations for a set of different volumes was fitted to the Birch--Murnaghan equation of state. For some of the presented systems, it is known that the cubic structure is not the global energy
minimum and that a tetragonal derivative exists.\cite{Gasi:2013fe,Liu:2005eh,Luo:2010bq,Balke:2007ebba} In this stage, these structural instabilities have been ignored deferring an in-depth discussion of these instabilities to an upcoming publication. 

\section{\label{sec:Results}Results}
\subsection{\label{subsec:SlatPaul}The Slater--Pauling Rule}
The results for the total magnetic moments are shown by means of an SP--curve, which is valid for many
of the known Heusler compounds. As seen in Fig.~\ref{fig1-1}, the
calculated magnetic moments for each compound in the series
Mn$_2$Y$^{\left(3d\right)}$Ga are well reproduced. For compounds
involving Al on the $Z$ position, the values similarly follow the SP
rule and are listed in Table~\ref{tab:Latdat}. Even for compounds formed
with valence electrons less than ${N_{\rm V}=24}$,  SP behavior is
observed, although with formally negative magnetic moments. The reason
for this is that the minority spin count for ${N_{\rm V}>24}$  is 12,
which remains true for ${N_{\rm V}<24}$, albeit with a change of the spin
channel, which becomes the majority spin count; this results in the
formally negative moments.
 
For every rule, there are exceptions. For compounds with late transition
metals as $Y$ atoms (for Ni and Cu, Pd and Ag, like Pt and Au), the
half-metallicity cannot be maintained. Mn$_2$NiAl and Mn$_2$NiGa are not
half-metallic; however, their magnetic moments are almost integers, and
thus these compounds could be mistaken for half-metallic
ferromagnets. Also, for Mn$_2$CuAl, Mn$_2$CuGa, Mn$_2$PdGa, Mn$_2$AgGa,
Mn$_2$PtGa, and Mn$_2$AuGa, the SP--behavior is no longer valid and the
expected moments are smaller than those predicted by the SP rule. The
moments decrease and assume non-integer values, as can be seen in
Fig.~\ref{fig1-1}. The reason is explained later in
paragraph~\ref{subsec:ElStruct} and Figs.~\ref{fig:DOS_Mn2YZ_216} and
\ref{fig:DOS_Mn2Y4dZ_216}.
\begin{table*}[htbp]
\begin{ruledtabular}
\begin{tabular}{lcccc|cccccc}
                        &$N_{\rm V}$  &       SG      &       $a_{\rm opt}$&      $M_{\rm opt}$               &       $a_{\rm exp}$&    $M_{\rm exp}$ &   Ref.   &       $a_{\rm theo}$      &$M_{\rm theo}$     &       Ref.  \\\hline   
Mn$_2$TiAl      &       21      &       225     &       5.96    &       -2.98   &               &                 &                                                  &       5.96    &-2.98&\cite{Meinert:2011ej}              \\              
Mn$_2$VAl       &       22      &       225     &       5.81    &       -2.00  &       5.92&   -1.94   &\cite{Buschow:1981hv},\cite{Jiang:2001kd} &       5.80        &       -2.00&\cite{Luo:2008fi}                                      \\   
Mn$_2$CrAl      &       23      &       225     &       5.73    &       -1.04  &               &                    &                                                  &       5.71    &       -1.00   &       \cite{Luo:2008fi}             \\  
Mn$_3$Al        &       24      &       225     &       5.80    &       0.00    &               &                    &                                                  &       5.80    &       0.00    &       \cite{Wurmehl:2006db}          \\ 
Mn$_2$FeAl      &       25      &       216     &       5.76    &       1.00    &               &                    &                                                  &       5.73    &       1.01    &       \cite{Luo:2008fi}              \\ 
Mn$_2$CoAl      &       26      &       216     &       5.74    &       2.00    &       5.80&   2.00    &       \cite{Ouardi:2013koba}  &       5.75                    &       2.00    &       \cite{Luo:2008fi}                               \\
Mn$_2$NiAl      &       27      &       216     &       5.81    &       1.19    &               &                    &                                                  &       5.64    &       1.44    &       \cite{Luo:2010bq}               \\
Mn$_2$CuAl      &       28      &       216     &       5.90    &       0.20    &       5.91&   0.22    & \cite{Feng:2010ec}   &       5.85                       &       0.22    &       \cite{Li:2009hy}                                 \\\hline
Mn$_2$TiGa      &       21      &       225     &       5.95    &       -2.97  &               &                    &                                                  &       5.95    &       2.98    &       \cite{Meinert:2011ej}           \\
Mn$_2$VGa       &       22      &       225     &       5.82    &       -1.98   &       6.10&   $-1.66$   &                   \cite{Buschow:1981hv}   &                           &                       &                                  \\     
Mn$_2$CrGa      &       23      &       225     &       5.76    &       -1.00   &               &                    &                                                  &       5.71    &       -1.00   &       \cite{Luo:2008fh}          \\     
Mn$_3$Ga        &       24      &       225     &       5.82    &       0.01    &               &                    &\cite{Balke:2007ebba}                             &       5.82    &       0.01    &       \cite{Wurmehl:2006db}                         \\    
Mn$_2$FeGa      &       25      &       216     &       5.76    &       1.03    &               &                    &       \cite{Gasi:2013fe}                         &       5.80    &       1.05    &       \cite{Luo:2008kb}           \\    
Mn$_2$CoGa      &       26      &       216     &       5.78    &       2.00    &               &                    &                                                  &       5.86    &                       &       \cite{Meinert:2011hn}\\   
Mn$_2$NiGa      &       27      &       216     &       5.85    &       1.18    &       5.91&      &        \cite{Liu:2006dx}   &                          &     1.28                &                                           \cite{Liu:2006dx} \\   
Mn$_2$CuGa      &       28      &       216     &       5.94    &       0.33    &               &                    &                                                  &       5.94    &       0.33    &       \cite{Chakrabarti:2013fd}    \\\hline
Mn$_2$ZrGa      &       21      &       225     &       6.14    &       -3.00   &               &                    &                                                  &                       &                       &                                           \\   
Mn$_2$NbGa      &       22      &       225     &       6.00    &       -2.00   &               &                    &                                                  &                       &                       &                                           \\   
Mn$_2$MoGa      &       23      &       225     &       5.91    &       -1.01   &               &                    &                                                  &                       &                       &                                           \\   
Mn$_2$RuGa      &       25      &       216     &       5.96    &       1.03    &       6.00         &   1.15             &                                   \cite{Hori:2002kzba}             &                       &             &                \\   
Mn$_2$RhGa      &       26      &       216     &       5.98    &       1.64    &               &                    &                                                  &                       &                       &                                          \\   
Mn$_2$PdGa      &       27      &       216     &       6.12    &       0.55    &               &                    &                                                  &                       &                       &                                          \\   
Mn$_2$AgGa      &       28      &       216     &       6.22    &       0.34    &               &                    &                                                  &                       &                       &                                          \\   
\end{tabular}
\caption{\label{tab:Latdat} Numerically optimized lattice parameters
  $a_{\rm opt}$ and magnetization $M_{\rm opt}$ of Mn$_2$-based Heusler
  alloys compared to the experimental ($a_{\rm exp}$, $M_{\rm exp}$) and
  theoretical ($a_{\rm theo}$, $M_{\rm theo}$) literature data.}
\end{ruledtabular}
\end{table*}

\subsection{\label{subsec:StrucMagGround}Structure and Magnetic Ground State}

The crystal structures for the Heusler compounds on the SP curve (SPC)
occur in two different modifications. For a valence count of ${N_{\rm V}\le24}$, the L2$_1$ (SG 255) type is favored, whereas the X$_a$ type (SG
216) is found for ${N_{\rm V}>24}$ (see also Table~\ref{tab:Latdat},
where the results of the structural relaxation are summarized next to
the known literature data for both experimental and theoretical studies
of cubic alloys, if available). Comparisons with known literature values
emphasize the agreement of our work with previous research
in the field. Minor discrepancies are usually related to either
deviations in the applied theoretical methods and details or to the
comparison of high-temperature measurements with zero-temperature
ground-state calculations. A disorder-induced decrease of the magnetic
moments is also possible.\cite{Hori:2002kzba} If the ground-state
lattice parameters differ, so do the resulting magnetic moments.

From the rule-of-thumb (Sec.\ref{sec:Cryst}) and other theoretical work\cite{Luo:2008fi} on a number of  examples, we know that, on the
one hand, Mn$_2$-based Heusler compounds with early transition elements
on the \textsl{Y} position adopt the L2$_1$-type structure. On the other
hand, compounds involving late transition metals \textsl{Y} adopt the
X$_a$-type inverse Heusler structure.  Table~\ref{tab:Latdat} reveals that this kind of behavior is true for all Mn$_2$-based Heusler
compounds involving $3d$ TMs on the \textsl{Y} position
(Mn$_2$\textsl{Y}$^{(3d)}$Ga) as well as  for the $4d$ transition metal
series (Mn$_2$\textsl{Y}$^{(4d)}$Ga).
 
Along with the structural transition, a magnetic transition from
parallel to anti-parallel alignment of Mn spins takes place due to Mn
occupying site \textsl{4b}. This begins at the transition point 
$N_{\rm V}^{\rm C}$, where the magnetization vanishes.\cite{Wurmehl:2006db} 
The type of ferrimagnetic ordering changes and becomes more complex, as will be pointed out in the following section.

\subsection{Stability Considerations and Heats of Formation}

\begin{figure}[htbp]
\begin{flushleft}
\includegraphics[width=.47\textwidth]{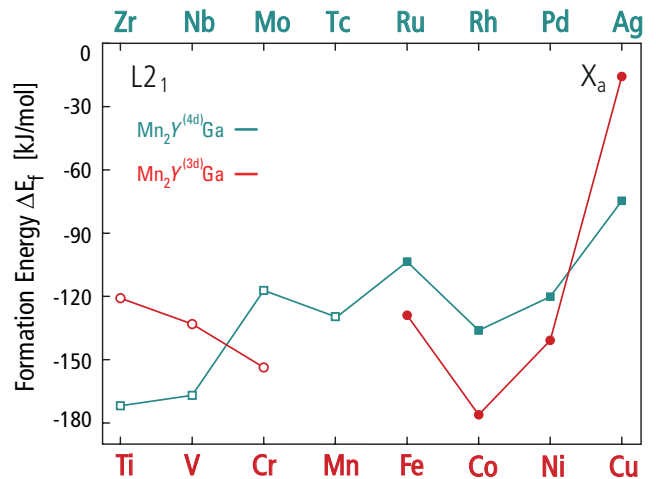}
\end{flushleft}
\caption{\label{fig:hof}Formation energies estimated at ${T=0}$ for  Mn$_2$\textsl{Y}$^{(3d)}$Ga and
  Mn$_2$\textsl{Y}$^{(4d)}$Ga compounds, as initial guess regarding the stability of these compounds. Open
symbols denote the L2$_1$-type structure. Filled symbols denote the X$_a$-type structure.}
\end{figure}

For estimating the stability of the investigated compounds, the heats of
formation were calculated from the elemental crystals in their
ground-state modification (except for Mn, where an anti-ferromagnetic
bcc structure was used as an approximation) using total energy
differences. Fig.~\ref{fig:hof} 
shows the changes in the approximated formation energy within the series of compounds under study. One sees
that the compounds are stable within the range of $-100$ to
$-180$~kJ/mol, except for Mn$_2$AgGa. 
The existence of some compounds such as Mn$_2$CoGa\cite{Li:2005gz} is
experimentally proven through synthesis. Nevertheless, phases
incorporating the investigated stoichiometries may exist in other
crystal structures, such as the tetragonal derivatives of cubic Heusler
structures. These findings will be presented in a later publication.

\begin{figure}[htbp]
\begin{flushleft}
\includegraphics[width=.47\textwidth]{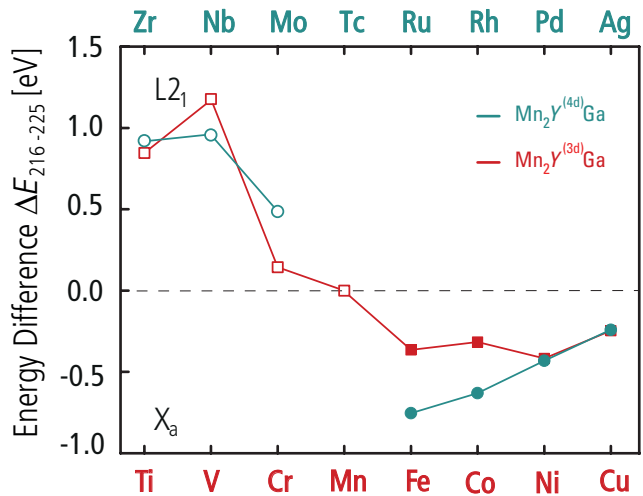}
\end{flushleft}
\caption{\label{fig:ediff} Total energy difference  between the inverse and regular 
structures ${\Delta E_{216-225} = E_{216}-E_{225}}$ calculated for
Mn$_2$\textsl{Y}$^{(3d)}$Ga and Mn$_2$\textsl{Y}$^{(4d)}$Ga cubic
Heusler materials.}
\end{figure}

Consideration of the total energy difference between the inverse and regular 
structures ${\Delta E_{216-225} = E_{216}-E_{225}}$ for the given
chemical composition \textsl{X}$_2$\textsl{YZ} (Fig.~\ref{fig:ediff}),
confirms that the \textsl{4b} site is going to be occupied by the
earlier transition metal among Mn and \textsl{Y}. 
This leads to the formation of the regular structures for the \textsl{Y}=Ti, V, Cr, Zr, Nb, Mo  
and inverse structures for the \textsl{Y}=Fe, Co, Ni, Cu, Ru, Rh, Pd and
Ag containing Heusler materials. 
The total energy difference can be interpreted as site-preference energy for the interchange of Mn and {\it Y} 
that distinguishes both structure types. For instance,  in Mn$_2$CrGa, with a rather small site-preference energy, 
indicating that the {\it regular} type structure is not strongly preferred, thermally induced disorder is more likely to occur than for example for Mn$_2$CoGa and Mn$_2$VGa, with solid energy differences.
Additional information about the stability of the phases under
investigation was obtained from the densities of state (DOS) that are
discussed in detail in the following section. It has proven useful to
look  for sharp peaks in the DOS at the Fermi edge that are symptoms of instabilities in the system. This  applies to
Mn$_2$FeGa\cite{Gasi:2013fe} and Mn$_3$Ga,\cite{Balke:2007ebba} which
have been experimentally realized in the tetragonal inverse structure.
It is known that besides Mn$_2$FeGa and Mn$_3$Ga there are  tetragonal phases for the $3d$ transition
metals (TMs)  involving Mn-based compounds, such as Mn$_3$Al, and Mn$_2$NiGa. Furthermore, we
could show that tetragonal alloys may exist in the series involving late
TMs of the IV period. A detailed discussion of these phases shall be
postponed to a later publication; we emphasise, however,
that possible  tetragonal derivatives from cubic Heusler phases
do not prove their existence, as other relaxation mechanisms such as
disorder phenomena may minimize the total energy. Nevertheless, the
instabilities do give a strong hint, as has been seen in {\it ab-initio}
studies related to shape-memory compounds.

\begin{figure}[htbp]
\includegraphics[width=0.45\textwidth]{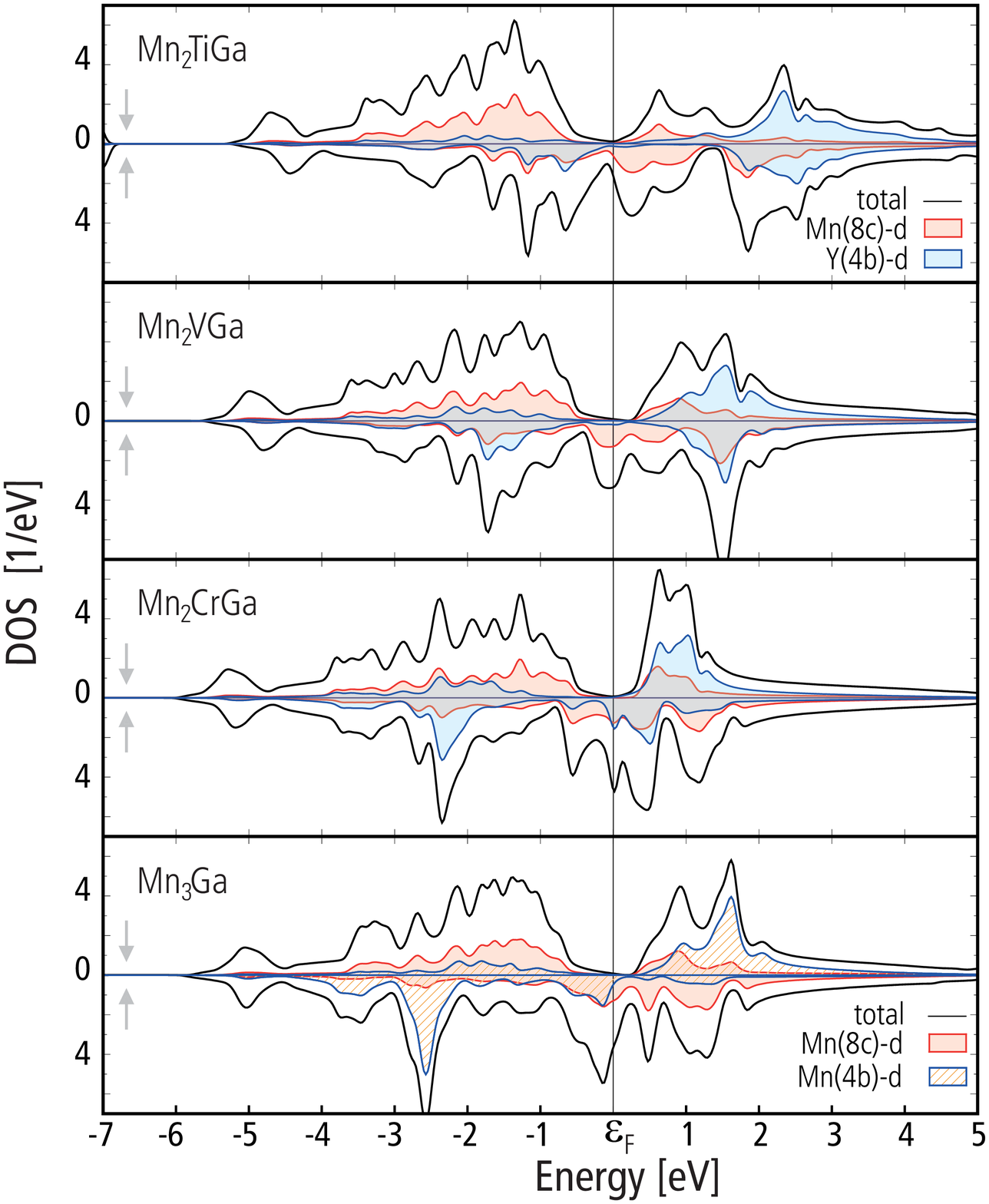}
\caption{\label{fig:DOS_Mn2YZ_225}Densities of states of the Mn$_2$\textsl{Y}$^{(3d)}$Ga compounds with \textsl{Y}=Ti, V, Cr, Mn.}
\end{figure}
\begin{figure}[htbp]
\includegraphics[width=0.45\textwidth]{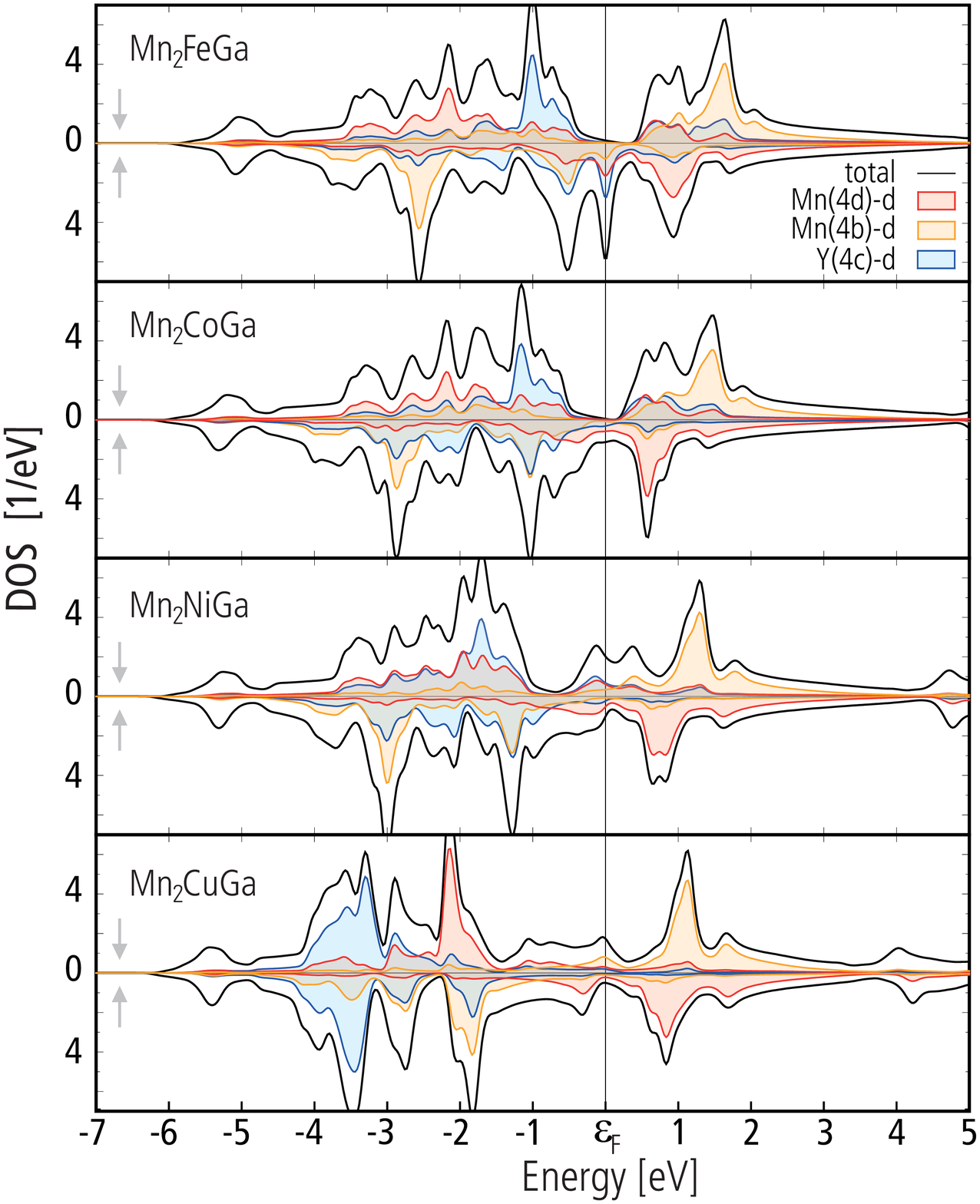}
\caption{\label{fig:DOS_Mn2YZ_216}Densities of states of the Mn$_2$\textsl{Y}$^{(3d)}$Ga compounds with \textsl{Y}=Fe, Co, Ni, Cu.}
\end{figure}
\begin{figure}[htbp]
\includegraphics[width=0.45\textwidth]{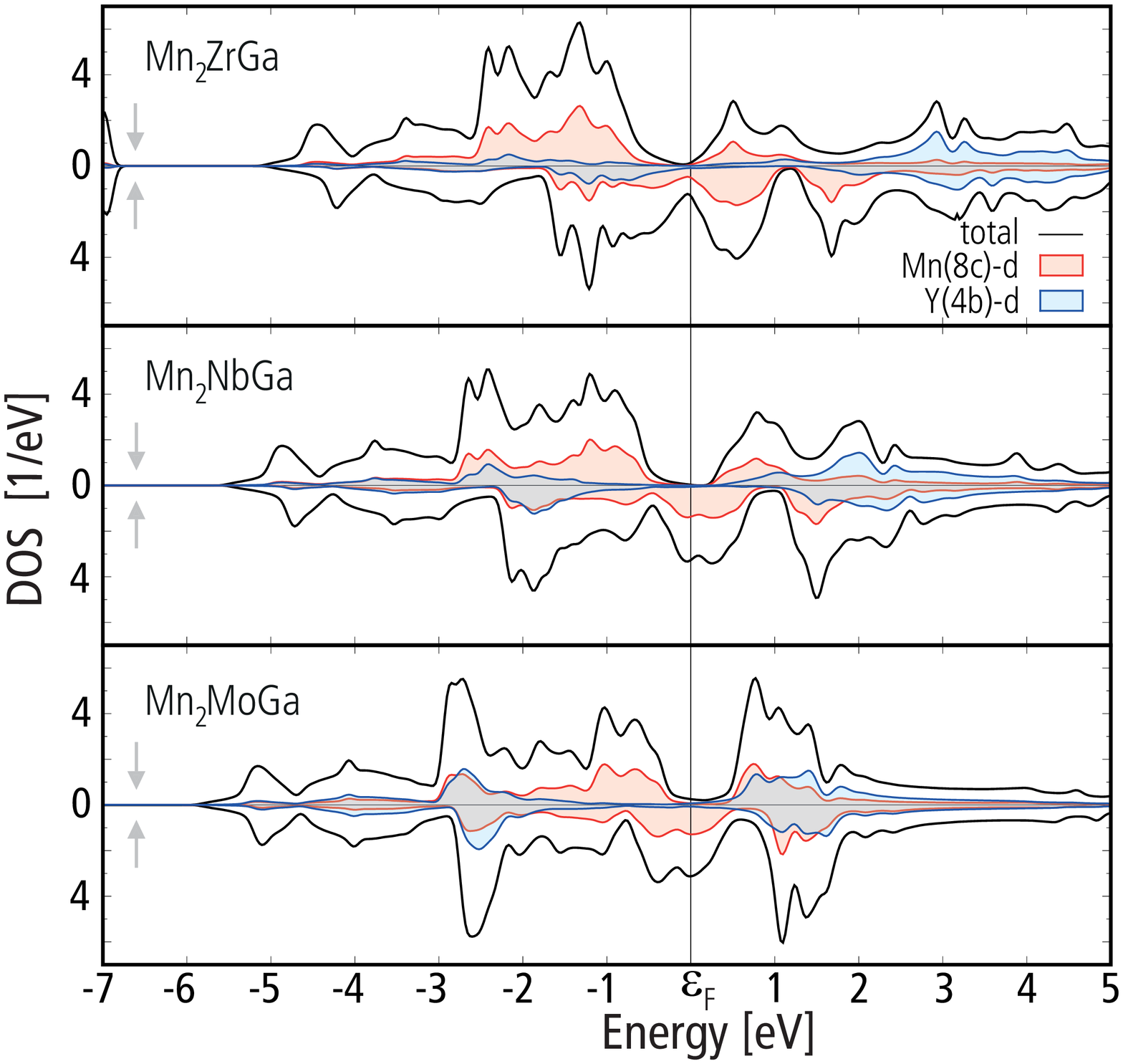}
\caption{\label{fig:DOS_Mn2Y4dZ_225}Densities of states of the Mn$_2$\textsl{Y}$^{(4d)}$Ga compounds with \textsl{Y}=Zr, Nb, Mo.}
\end{figure}
\begin{figure}[htbp]
\includegraphics[width=0.45\textwidth]{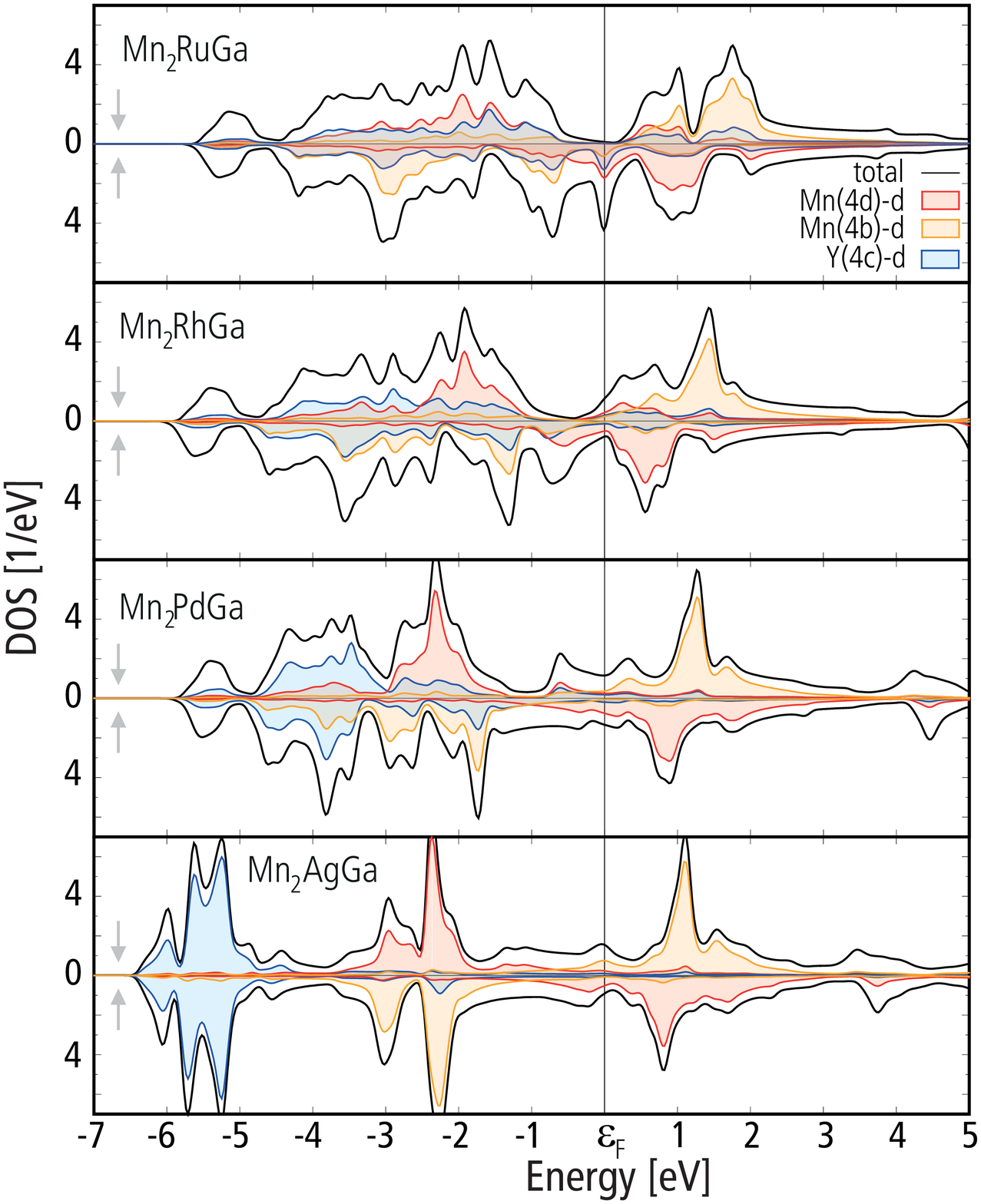}
\caption{\label{fig:DOS_Mn2Y4dZ_216}Densities of states of the Mn$_2$\textsl{Y}$^{(4d)}$Ga compounds with \textsl{Y}=Ru, Rh, Pd, Ag.}
\end{figure}    

\subsection{\label{subsec:ElStruct}Electronic Structure}

The calculations reveal that nearly all compounds are at least pseudo-half-metallic or even truly half-metallic with the associated gap in the \textit{minority} densities of state (DOS) for compounds with a valence electron count of less than 24 and a gap in the \textit{majority} DOS for more than 24 electrons. It is worth mentioning that the definition of spin-up and spin-down electrons is connected with the choice of the magnetic moment as positive. However, we may allow negative values to emphasize the occupancy of the $d$--states  according to the SP rule, although the notion of negative moments is unphysical. Still doing so leads to a better understanding of the mechanism of magnetic ordering. Although Figs. \ref{fig:DOS_Mn2YZ_225} and \ref{fig:DOS_Mn2YZ_216} were chosen to exhibit the gap in the spin-down channel for all compounds, the minority channel changes from the spin-down state to the spin-up state as the electron count crosses 24 electrons per formula unit.

\begin{table}[htbp]
\begin{tabular}{cc|lc|lc}
$N_{\rm V}$ & $N_{{\rm V},\textsl{Y}}$ & Mn$_2$\textsl{Y}$^{(3d)}$Ga & \multicolumn{1}{l}{$P$/\%} & Mn$_2$\textsl{Y}$^{(4d)}$Ga & \multicolumn{1}{l}{$P$/\%} \\ \hline
21 & 4& Mn$_2$TiGa & 83 &  Mn$_2$ZrGa & 82 \\ 
22 & 5& Mn$_2$VGa & 94 &   Mn$_2$NbGa & 98 \\ 
23 & 6& Mn$_2$CrGa & 97 &  Mn$_2$MoGa & 85 \\ 
24 & 7& Mn$_3$Ga & 96 &  &\\   
25 & 8& Mn$_2$FeGa & 95 &  Mn$_2$RuGa & 95 \\ 
26 & 9& Mn$_2$CoGa & 93 &  Mn$_2$RhGa & 15 \\ 
27 & 10&        Mn$_2$NiGa & 35 &  Mn$_2$PdGa & 7 \\  
28 & 11&        Mn$_2$CuGa & 53 &  Mn$_2$AgGa & 24 \\\hline
\end{tabular}
\caption{\label{tab:SpinPol}The total number of the valence electrons,
  $N_{\rm V}$ and valence electrons of the \textsl{Y} atom, $N_{\rm V,{\textsl{Y}}}$, and the calculated spin polarization, \textsl{P},
  of the DOS at the Fermi edge of the  Mn$_2$\textsl{Y}$^{(3d)}$Ga and
  Mn$_2$\textsl{Y}$^{(4d)}$Ga series.}
\end{table}

In Figs. \ref{fig:DOS_Mn2YZ_225} and \ref{fig:DOS_Mn2YZ_216}, the DOS for both structure types are compared. A
feature seen in the DOS is that the gap at the Fermi energy is bordered
by Mn states for the L2$_1$-type structure, whereas this is not the case
for the inverse compounds. A second property is the prominent downward
trend of the $Y$ $d$--states. Independent of the structure type, the
local environment of Mn exhibits tetrahedral  symmetry, which leads
to the SP curve being continuous across the transition point at $N_{\rm
  V}=24$, leaving the near-half metallic behavior unchanged. Following
Ouardi and Galankis, \cite{Ouardi:2011bt,Galanakis:2002fk} the gap in
the minority channel in Mn$_2$-based alloys is the result of
hybridization of the $X$ atoms on site \textsl{8c} in the SG
225, which are again tetrahedrally coordinated with their neighboring
atoms. The final size of the gap is determined by the crystal field
splitting of the $e_{g}$ and $t_{2g}$ states (at the $\Gamma$ point),
which is determined by the symmetry and coordination of the Mn atoms.
 
The DOS for Mn$_2$NiGa and Mn$_2$CuGa explains the deviation from the SP curve because the gap at the Fermi energy closes. This is so because the $d$-electron states of Ni and Cu disappear from the Fermi energy being drawn into the core. The states at the Fermi edge are now mainly composed of $s$ electrons for Mn$_2$AgGa. Condensing this finding into one number that is closely connected with the concept of half-metallicity, the spin polarization, $P$, emphasizes the preceding facts and is listed for the investigated compounds in Table \ref{tab:SpinPol}. Once more, the failure of the SP rule for compounds with more than 26 valence electrons is apparent in terms of the spin polarization.

\subsection{\label{subsec:LocMom}Local Magnetic Moments}

A truly large  number of publications on the SP behavior of various compounds exists,\cite{Skaftouros:2013ftba,  Kubler:2007im, Galanakis:2005vz} and
yet the local magnetic structure has not been intensively investigated so far, even though the local 
magnetic structure is closely related with the magnetic phenomenon. 
First-principles calculations are a sophisticated tool to access site-resolved quantities such as local magnetic 
moments, even though the partitioning of a solid into atomic regions is an arbitrary procedure
and approximate projection techniques need to be applied in most cases. 
But as experimental information on atomic magnetic moments in bulk Heusler materials is
rather incomplete, we focus our attention on the local magnetic and electronic environment 
of the Mn atoms in both structure types. We show that the magnetic moment in inverse 
Heusler compounds (X$_a$) of one Mn atom is locked at about $3~\mu_{\rm B}$, whereas no such rule is found for the local moments of the Mn-based L2$_1$-type compounds, where 
 Mn  occupies the tetrahedrally coordinated position \textsl{8c}.

\begin{figure*}[htbp]
\includegraphics[width=\textwidth]{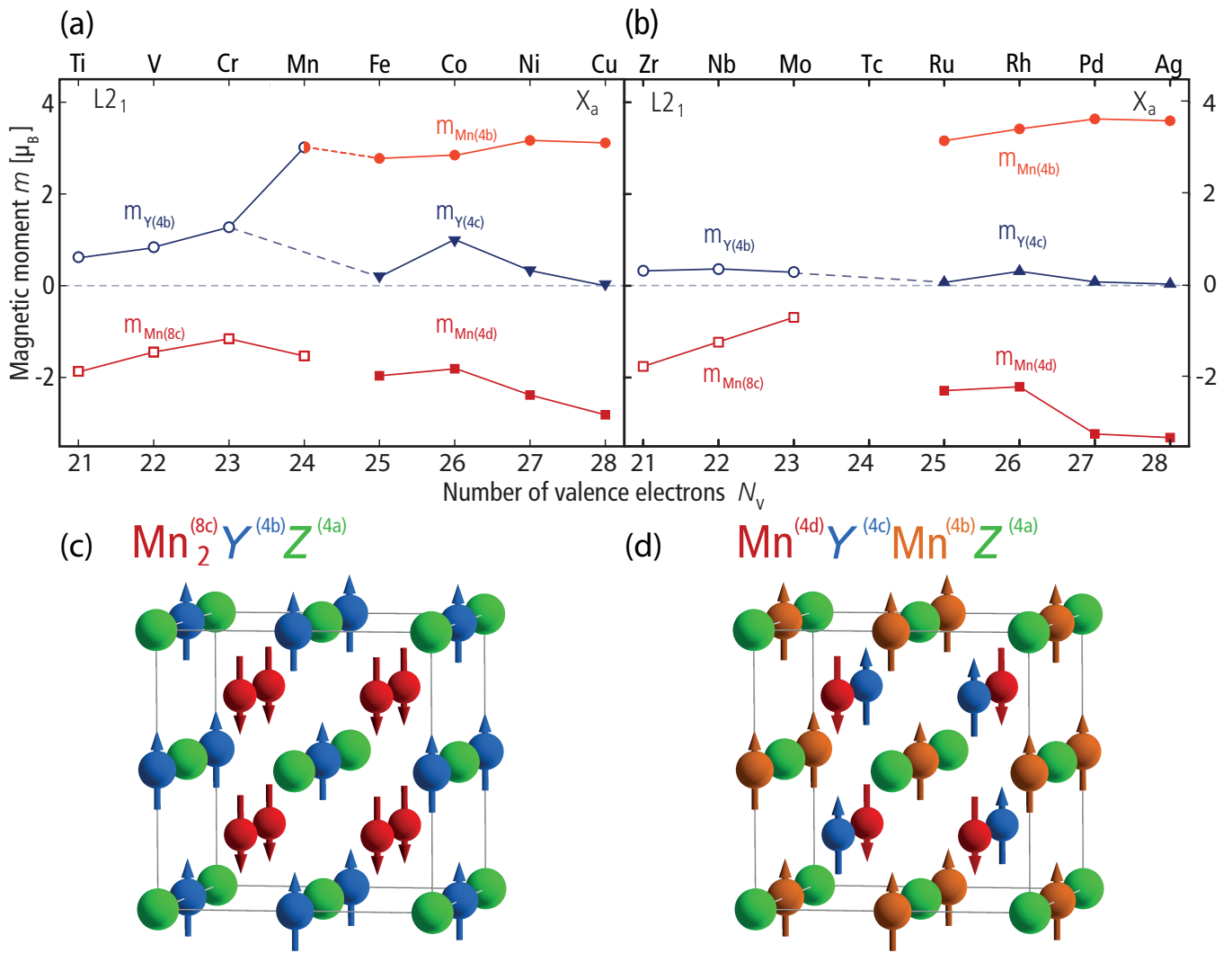}
\caption{\label{fig:LocMomMn2Y34dGa}Atomic magnetic moments  in (a)~Mn$_2$\textsl{Y}$^{(3d)}$Ga and
  (b)~Mn$_2$\textsl{Y}$^{(4d)}$Ga-compounds. Open symbols denote the L2$_1$-type structure. Filled symbols denote the X$_a$-type structure. Panels~(c) and (d) show the
  magnetic ordering corresponding to the L2$_1$ and X$_{a}$ types, respectively.  }
\end{figure*}

To see this, the local  magnetic moments  of the Mn$_2$\textsl{Y}$^{(3d)}$Ga and Mn$_2$\textsl{Y}$^{(4d)}$Ga 
compounds  are shown in Fig.~\ref{fig:LocMomMn2Y34dGa}, which in panels c and d provides a schematic overview of the type of magnetic order. The left-hand side of the plots
 Fig.~\ref{fig:LocMomMn2Y34dGa}~(a, b)  for ${N_{\rm V}<24}$ regime  displays the local 
 moments of the transition metal atoms in the L2$_1$ structure. The electron count  increases through the variation of the \textsl{Y} atom 
 occupation from \textsl{Y}=Ti to Cu and from \textsl{Y}=Zr to Ag. As the electron count is increased by one, the magnetic 
 moments of the systems, following the SP curve, increase by $1~\mu_{\rm B}$. Considering the series Mn$_2$\textsl{Y}$^{(3d)}$Ga,
  we recognize that the moments of \textsl{Y}(\textsl{4b}) and
  Mn(\textsl{8c}) display almost linear behavior. On the other hand, as the 
  magnetic moment on \textsl{Y}(\textsl{4b}) increases, the absolute
  value of Mn(\textsl{8c}) decreases for both series  with a change of slope at the Cr position. Increasing the electron count therefore leads to a filling of either the minority channel 
  (${N_{\rm V}\leq24}$) or the majority channel (${N_{\rm V}\ge24}$).
\begin{table}[htbp]
\caption{Atomic magnetic moments in Mn$_2$\textsl{Y}$^{(3d)}$Ga Heusler
  compounds; $M$ is the total magnetization.}
\label{tab:MagMomMn2Y3dGa}
\begin{tabular}{l|cc|cccc|c}
\hline
regular & $N_{\rm V}$ &  $a_{\rm opt}$ & Mn(\textsl{8c}) & $-$ & \textsl{Y}(\textsl{4b}) & \textsl{Z}(\textsl{4a}) & $M$ \\ \hline
Mn$_2$TiGa & 21 &  5.95 & -1.87 & &0.62 & 0.05 & -2.97 \\ 
Mn$_2$VGa & 22 &  5.82 & -1.45 & &0.84 & 0.02 & -1.98 \\ 
Mn$_2$CrGa & 23  & 5.76 & -1.16 && 1.27 & 0.00 & -1.00 \\ 
Mn$_3$Ga & 24  & 5.82 & -1.53 & &3.02 & 0.04 & 0.01 \\ \hline
inverse & $N_{\rm V}$   & $a_{\rm opt}$ & Mn(\textsl{4d}) & \textsl{Y}(\textsl{4c}) & Mn(\textsl{4b}) & \textsl{Z}(\textsl{4a}) & $M$ \\ \hline
Mn$_2$FeGa & 25 &  5.76 & -1.97 & 0.21 & 2.78 & 0.01 & 1.03 \\ 
Mn$_2$CoGa & 26 &  5.78 & -1.81 & 1.00 & 2.85 & -0.01 & 2.00 \\ 
Mn$_2$NiGa & 27 &  5.85 & -2.38 & 0.34 & 3.17 & 0.01 & 1.18 \\ 
Mn$_2$CuGa & 28 &  5.94 & -2.81 & 0.04 & 3.11 & 0.00 & 0.33 \\ \hline
\end{tabular}
\end{table}

\begin{table}[htbp]
\caption{Atomic magnetic moments in Mn$_2$\textsl{Y}$^{(4d)}$Ga
  Heusler compounds; $M$ is the total magnetization.}
\label{tab:MagMomMn2Y4dGa}
\begin{tabular}{l|cc|cccc|c}\hline
regular & $N_{\rm V}$ &  $a_{\rm opt}$ & Mn(\textsl{8c}) & $-$ & \textsl{Y}(\textsl{4b}) & \textsl{Z}(\textsl{4a}) & $M$ \\ \hline
Mn$_2$ZrGa & 21 &  6.14 & -1.76 &  & 0.33 & 0.07 & -3.00 \\ 
Mn$_2$NbGa &22 &  6.00 & -1.23 &  & 0.36 & 0.03 & -2.00 \\ 
Mn$_2$MoGa &23 &  5.91 & -0.69 &  & 0.30 & 0.02 & -1.01 \\ \hline
inverse & $N_{\rm V}$   & $a_{\rm opt}$ & Mn(\textsl{4d}) & \textsl{Y}(\textsl{4c}) & Mn(\textsl{4b}) & \textsl{Z}(\textsl{4a}) & $M$ \\ \hline
Mn$_2$RuGa& 25 &  5.96 & -2.29 & 0.07 & 3.16 & 0.03 & 1.03 \\ 
Mn$_2$RhGa& 26 &   5.98 & -2.20 & 0.31 & 3.42 & 0.03 & 1.64 \\ 
Mn$_2$PdGa& 27 & 6.12 & -3.23 & 0.08 & 3.64 & 0.02 & 0.55 \\ 
Mn$_2$AgGa&28 &  6.22 & -3.31 & 0.04 & 3.60 & 0.01 & 0.34 \\ \hline
\end{tabular}
\end{table}

As shown in section~\ref{subsec:ElStruct}, every compound is at least nearly half-metallic, reflecting the change of the moments 
by means of the site-resolved densities of state (Fig.~\ref{fig:DOS_Mn2YZ_225}). The decreasing number of states in the 
minority channel and the increasing number of states in the majority channel for both Mn and \textsl{Y} atoms are related to the change 
in the local moments. Filling states of \textsl{Y} and Mn atoms at the same time, where the Mn magnetic moment is 
aligned anti-parallel to the \textsl{Y} atom and is consequently decreasing. The trends of the L2$_1$ Mn-based Heusler compounds intriguingly resemble 
the nature of the local moments in Co$_2$-based Heusler
compounds\label{ref:Co2Heus}.\cite{Kandpal:2007tn}

Moving forward in the series,  perhaps the most outstanding Heusler alloy, Mn$_3$Ga, marks the transition point between both 
cubic structure types. Although Mn$_3$Ga is found to adopt the L2$_1$ structure in the cubic approximation, it exhibits properties 
of the inverse Heusler structure, namely, the anti-parallel alignment of Mn atoms on sites \textsl{4d} and \textsl{4b}. The 
determining characteristic property of Mn-based inverse Heusler compounds clearly is this antiparallel alignment. This is displayed 
in Fig.~\ref{fig:LocMomMn2Y34dGa}. Through its peculiar behavior, Mn$_3$Ga is not only  an exceptional compound of the L2$_1$ 
series, but it also constitutes the \textsl{transition point} between both structure types. Additionally, the total magnetic moment for 
Mn$_3$Ga is supposed to vanish,\cite{Wurmehl:2006db} as it marks the \textsl{compensation point}, following the SP 
rule as well. It is emphasized again that tetragonal Heusler compounds were not considered for this part of the study, and it is
 noted again that Mn$_3$Ga does not exhibit a cubic structure; as with Mn$_3$Al, Mn$_2$FeGa, Mn$_2$NiGa, and Mn$_2$NiAl,  Mn$_3$Ga has
 been found to be tetragonal.

 With  Mn occupying  the \textsl{4b} site, the local moment on site $4d$ increases in 
 absolute value. But most important, the local magnetic moment of Mn on site \textsl{4b} is locked at $3~\mu_{\rm B}$. Although 
 some compounds are known in the literature, a comprehensive overview of the local magnetic structure has not been provided yet. 
 Therefore, the remarkable consistency of the moment on site \textsl{4b} has not been appreciated enough.

Beyond the transition point, the magnetic ordering, following the crystallographic ordering, changes from parallel to antiparallel alignment
of the Mn spins. Owing to a symmetry reduction of index t2 (\textit{trans\-la\-tionen\-gleiche} subgroup), the Wyckoff position \textsl{8c}
splits into \textsl{4d} and \textsl{4c}, which are now occupied by Mn(\textsl{4d}) and {\it Y}(\textsl{4c}), the second Mn 
atom now occupying site \textsl{4b}, which is electronically distinct from site \textsl{4d}. The electric potential experienced on the Mn 
sites is different owing to the nearest neighbors, i.e. Mn(\textsl{4d})
is coordinated by four Ga(\textsl{4a}) and four Mn(\textsl{4b}) 
atoms, whereas  Mn(\textsl{4b}) is surrounded by four Mn(\textsl{4d}) and
four {\textsl Y}(\textsl{4c}) atoms. One observes that the 
nearest-neighbor combination, Mn(\textsl{4b})\,--{\textsl Y}(\textsl{4c}), results in a high local moment on Mn(\textsl{4b}) 
and ferromagnetic coupling of both species, whereas the
Mn(\textsl{4b})\,--Mn(\textsl{4d}) coupling is antiferromagnetic, as in
the elemental crystal.\label{antiferroMn}

While varying the {\textsl Y}$^{(4d)}$ atoms, the {\textsl Y} atom $d$ states decrease in energy and act as charge sinks, leading to an 
increase of the magnetic moment on the Mn(\textsl{4d}) position owing to its atomic states and chemical environment. The character 
of the  {\textsl Y} atom $d$ state is therefore reflected in the energy levels of the Heusler compounds. This behavior of the magnetic 
moments with increasing electron count is demonstrated in
Fig.~\ref{fig:LocMomMn2Y34dGa}.

The compounds shown on the right-hand side of Figures~\ref{fig:LocMomMn2Y34dGa}~(a,\,b) are composed by the data for the 
inverse compounds, and thus the contributions to the change of the total magnetic moment stem from the Mn atoms on position \textsl{4d} only. Meanwhile, the change in the local moments of the 
\textsl{Y} atoms is significant for Co only, whereas Fe, Ni, and Cu carry small moments  that do not follow a particular trend.

Furthermore, the Mn-based Heusler compounds incorporating transition metals from the IV period are studied in the same fashion. 
In contrast to the  series involving $3d$ transition metals, the magnetic moments are largely built by the Mn atoms only. The 
contribution of the transition metals is of minor importance. As for the lighter homologues, the SP behavior breaks down for 
alloys with 27 or more valence electrons. The local magnetic moments almost compensate, as the transition metals on position 
\textsl{4c} (Pd,Ag) do not contribute to the total moment.

\subsection{\label{subsec:Exchge}Exchange Coupling and Curie
  Temperatures}

The interaction of the magnetic moments is typically parametrized using
the effective Heisenberg Hamiltonian, where the interaction strength is
described by the so-called exchange constants or pair interaction 
energies $J_{ij}$:  $H=-\sum_{ij}J_{ij}\hat{e}_i\hat{e}_j$, where
$\hat{e}_{i,j}$ are the directional unit vectors along the magnetic
moments on sites $i$ and $j$. Knowing the set of $\{J_{ij}\}$, one can estimate
the magnetic order (positive $J$ corresponds to the parallel coupling,
negative - to the antiparallel one) and the ordering temperature (we will call it the Curie temperature
$T_{\rm C}$ also in case of a ferrimagnet). The proper way to calculate
$T_{\rm C}$ is via the exact $T$-dependent solutions of the parametrized Heisenberg model using Monte-Carlo 
simulations. In the present study, we are more interested in general
trends, rather than in exact solutions; therefore we will use the standard mean-field approximation\cite{Liechtenstein:1987br}
(MFA), for which ${k_{\rm B}T_{\rm C}=2/3\cdot J_{\rm max}}$, where $J_{\rm max}$ is the maximal eigenvalue of the
$\{J^{\mu\nu}_0\}$ matrix, with ${J^{\mu\nu}_0=\sum_{j\in\{\nu\}}J_{0j}}$, where 0 and $j$ are the site
indices, 0 - is fixed within the $\mu$ sublattice, $j$ - runs over the
$\nu$ sublattice. In this case, $J^{\mu\nu}_0$ represents the effective interaction
of the 0-th site from the sublattice $\mu$ with the whole sublattice
$\nu$. The sum over $\nu$ sites is supposed to be infinite;  in
practice it is truncated at a sufficiently large cluster radius, typically a few lattice constants.  
For the  calculation of $J_{ij}$  pair interactions  we  use the  real-space
Liechtenstein approach,\cite{Liechtenstein:1987br} which assumes thathe
magnetic moments and the band structure do not change when the directions of the moments are varied. These
calculations are carried out using the Munich SPR-KKR Green's
function based method.\cite{EKM11} Combined with the MFA,  we usually obtain a slightly overestimated $T_{\rm C}$.  

In
addition, we will use another complimentary formalism, the so-called spherical approximation (SPA)
proposed by Moriya.\cite{Mor85} In the SPA the exchange interactions are obtained
in reciprocal space as $J_{\nu \mu}(\bf k)$, where $\nu$ and $\mu$ designate sublattice-types. For each $\bf k$ in the Brillouin zone all eigenvalues of the matrix $J_{\nu \mu}(\bf k)$ are obtained, calling them $ j_{n,\bf k}$. Then $T_{\rm C}$ is  estimated  by means of
\begin{equation*}
 k_{\rm  B}T_{\rm C}=\frac{2}{3}\ M^2_{\rm tot}\Big[\sum_{\rm n,{\bf k}}
  j^{-1}_{n,{\bf k}}\Big]^{-1}, 
\end{equation*}  
  where $M^2_{\rm tot}$ is the sum of the
squared local moments.\cite{Kubler:2007im,Kubler:2009hz} In contrast to
the Liechtenstein formalism, this approach  accounts for the change of
the size of the magnetic moments upon their  rotations. In practice it is found to give slightly underestimated values of $T_{\rm C}$. The list of
calculated $T_{\rm C}$ is summarized in Table~\ref{tab:TC}.
\begin{table}[htbp]
\begin{ruledtabular}
\caption{\label{tab:TC}The Curie temperatures of Mn$_2$$Y$$^{\left(3d\right)}$Ga, Mn$_2$$Y$$^{\left(3d\right)}$Al
  and Mn$_2$$Y$$^{\left(4d\right)}$Ga Heusler compounds. Present
  calulations using MFA and SPA formalisms are referred  as $T_{\rm C,MFA}$
  and  $T_{\rm C,SPA}$; the values from the literature $T_{\rm C,lit}$
  are marked by the subscripts ``$t$'' or ``$e$'', which refer to theoretical
  calculations and experimental measurements, respectively.}
\begin{tabular}{llccc}
Material & $T_{\rm C,MFA}$~[K] & $T_{\rm C,SPA}$~[K] & $T_{\rm C,lit}$~[K] & Ref. \\ \hline
Mn$_2$TiGa & 557 & 525 & 663$_{\rm t}$ & \cite{Meinert:2011ej}\\ 
Mn$_2$VGa & 587 & 387 & 783$_{\rm e}$ & \cite{Kumar:2010gp} \\ 
Mn$_2$CrGa & 578 & 213 &  &  \\ 
Mn$_3$Ga & 221 & 314 & 482$_{\rm t}$ & \cite{Kubler:2006be} \\ 
Mn$_2$FeGa & 601 & 322 &  &  \\ 
Mn$_2$CoGa & 928 & 668 & 710$_{\rm e}$,886$_{\rm t}$ & \cite{Li:2005gz},\cite{Meinert:2011hn} \\ 
Mn$_2$NiGa & 1005 & 586 & 600$_{\rm e}$ & \cite{Blum:2011kg} \\ 
Mn$_2$CuGa & 1491 & 954 &  &  \\  \hline
Mn$_2$ZrGa & 207 & 185 &  &    \\ 
Mn$_2$NbGa & 289 & 239 &  &    \\ 
Mn$_2$MoGa & 140 & 81 &  &    \\ 
Mn$_2$RuGa & 619 & 369 & 560$_{\rm e}$ & \cite{Hori:2002kzba}  \\ 
Mn$_2$RhGa & 576 & 408 &  &    \\ 
Mn$_2$PdGa & 809 & 490 &  &    \\ 
Mn$_2$AgGa & 1240 & 751 &  &    \\ \hline
Mn$_2$TiAl & 428 &  & 665 & \cite{Meinert:2011ej}  \\ 
Mn$_2$VAl & 588 &  & 663$_{\rm t}$,638$_{\rm t}$,760$_{\rm e}$ &\cite{Kubler:2006be,Sasioglu:2005ex,Kumar:2010gp} \\ 
Mn$_2$CrAl & 549 &  &  &   \\ 
Mn$_3$Al & 306 & 342 &  &   \\ 
Mn$_2$FeAl & 614 & ~200 &  &    \\ 
Mn$_2$CoAl & 985 & 740 &  720$_{\rm e}$,890$_{\rm t}$ & \cite{Ouardi:2013koba,Meinert:2011hn}  \\ 
Mn$_2$NiAl & 1140 & 452 &  &    \\ 
Mn$_2$CuAl & 1539 & 884 &  &    \\ 
\end{tabular}
\end{ruledtabular}
\end{table}

\begin{figure*}[htbp]
        \begin{center}
                \includegraphics[width=\textwidth]{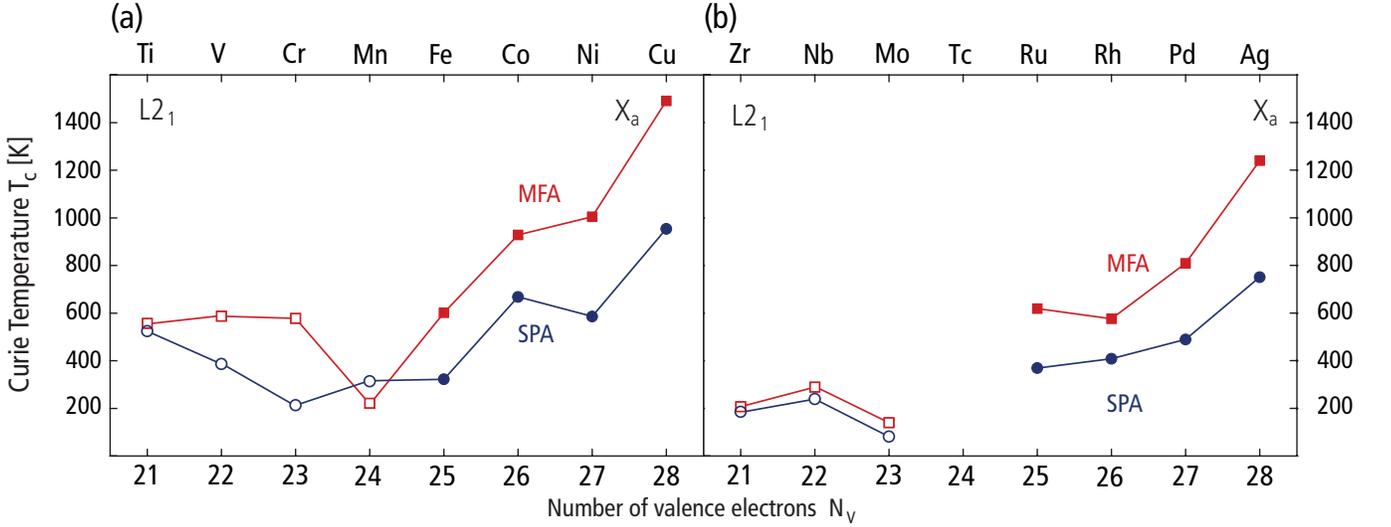}
                \caption{\label{fig:Tc}The calculated Curie temperatures of the Heusler compounds containing Ga. The results of the 
                        mean-field approximation (MFA) is compared with that of the spherical approximation (SPA). Open symbols: L2$_1$-type structure. Filled symbols:  X$_a$-type structure.}
        \end{center}
\end{figure*}

\begin{figure*}[htbp]
\includegraphics[width=\textwidth]{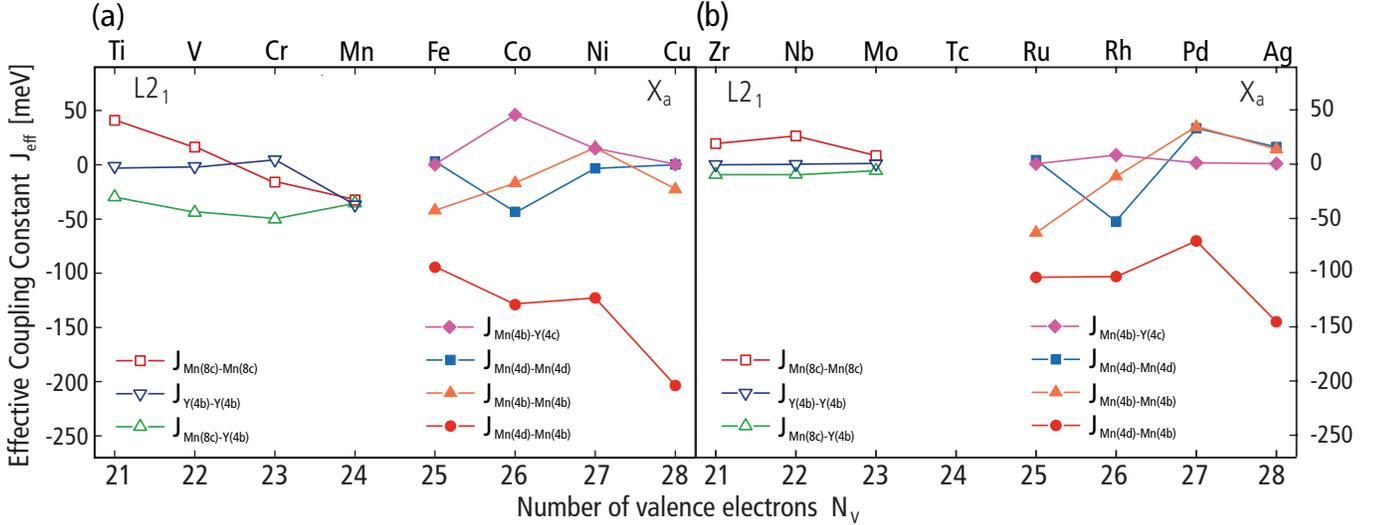}
\caption{\label{fig:Jeff-MnY34Ga} Effective exchange coupling constants $J^{\mu\nu}_{\rm 0}=\sum_{j\in\nu}J_{0j}$ for the Mn$_2$\textsl{Y}$^{(3d)}$Ga and Mn$_2$\textsl{Y}$^{(4d)}$Ga compounds. Open symbols: L2$_1$-type structure. Filled symbols:  X$_a$-type structure.}
\end{figure*}

Since the differences between magnetic moments in isoelectronic Mn$_2$\textsl{YZ} with \textsl{Z}=Al and Ga are rather small (Tab.~\ref{tab:TC}), we will restrict further discussion to the
set of Mn$_2Y$Ga compounds only.  Both MFA and SPA estimations of the Curie temperatures are compared in Fig.~\ref{fig:Tc}.
It follows, despite the noticeable difference, the general trends shown by both approaches are rather similar: with increasing
$N_{\rm V}$ the $T_{\rm C}$  decreases (very slightly in case of MFA and
more noticeably in case of SPA for \textsl{Y}$^{(3d)}$)  for the regular
compounds (${N_{\rm V}<24}$), and then significantly increases for the
inverse compounds (${N_{\rm V}>24}$). Such behavior to a large extent is
defined by the nearest (between the \textsl{4b} and \textsl{4c/4d}
sites) and next-nearest neighbor exchange coupling constants. Here we do
not plot them explicitly, but since the dominant contribution is contained  in 
the effective sublattice coupling $J^{\mu\nu}_{0}$ Figure~\ref{fig:Jeff-MnY34Ga}
gives an idea how the magnetic order is formed.
For ${N_{\rm V}<24}$ regime the nearest neighbors of Mn which sits in \textsl{8c} (\textsl{4c/4d}) are the weakly- or nonmagnetic elements in \textsl{4b}, i.e. \textsl{Y}=Ti, V and Cr. The corresponding
nearest-neighbor exchange coupling constants (estimated from the Liechtenstein approach) are
${J\approx}$~$-7$, $-9.8$ and $-11.7$~meV, respectively. Despite that most of the
Mn-Mn next-neighbor interactions (\textsl{4c-4d}, \textsl{4d-4d}, etc.) 
are also antiparallel, the dominating nearest \textsl{Y}\,--Mn antiparallel 
interaction typically fixes the ferromagnetic order in Mn(\textsl{8c})
sublattice, set by the indirect Mn(\textsl{8c})\,--\textsl{Y}(\textsl{4b})\,--Mn(\textsl{8c}) coupling.\cite{Sasioglu:2005ex}
Thus, the relatively low Curie temperatures for ${N_{\rm V}<24}$ can
be understood as a result of the relatively strong degree of magnetic frustration (i.e., the presence
of the competing interactions). This becomes especially evident in case
of Mn$_2$CrGa (${N_{\rm V}=23}$) and Mn$_3$Ga (${N_{\rm V}=24}$) which show the lowest ${T_{\rm
    C}\approx200}$~K within both the SPA and the MFA. The fact, that magnetic
frustration is just another side of electronic instability is clearly seen
in the density of states, which exhibits peaks at the
Fermi energy for both compounds. Furthermore, in spite of the
frustration   in the group of inverse compounds ${(N_{\rm V}>24)}$ 
being  lower, there are still a few unstable materials, such as Mn$_2$FeGa and
Mn$_2$RuGa: the first one is known to be tetragonal,\cite{Winterlik:2012cp} the second one relaxes through
chemical disorder.\cite{Hori:2002kzba} By increasing the number of valence electrons in the ${N_{\rm V}>24}$ regime
the cubic structure becomes more stable. Thus,   $T_{\rm C}$ increases for Co, Ni and Cu containing compounds. An important aspect within the inverse
group  is that the exchange coupling constants between
nearest-neighbors, i.e. between Mn in \textsl{4b} and Mn in \textsl{4c}, become very large:
$-20.1$, $-29.7$, $-33.2$ and $-40.2$ meV for Fe, Co, Ni and Cu containing
materials, respectively.  As we see, a similar situation
is observed for $Y^{(4d)}$ containing group as well.

\section{\label{sec:Conc}Summary}
The SP curves show that the magnetic moments of Mn-based 
compounds, independent of the structure type, are continuous across the compensation point and follow the 
SP rule (with the exception of compounds with ${N_{\rm V}\geq27}$). The total magnetization $M$ is composed 
of contributions of different characters (localized or itinerant) for the different structure types (Figs.~\ref{fig:LocMomMn2Y34dGa} 
and \ref{fig:Jeff-MnY34Ga}). Considering the empirical rules regarding the chemical ordering (Sec.~\ref{sec:Cryst}) and accounting for
the trends in the local moments, we can understand how the magnetization in Mn$_2${\it YZ} systems is formed. The L2$_1$-type 
Heusler compounds are purely itinerant ferrimagnets with small magnetic moments on the Mn atoms that are coupled
 ferromagnetically. The \textsl{Y} atoms on \textsl{4b} sites are of minor importance in the sum of the various contributions to the total moment. 
 They are coupled antiparallel to the Mn moments on the \textsl{8c} sites. The magnetism in the X$_a$-type (inverse) Heusler 
 compounds is composed of a large, localized\cite{Kubler:1983fj,Meinert:2011jqba}  system-independent moment on Mn(\textsl{4b})
 of about $3~\mu_{\rm B}$. The magnetic moment of Mn(\textsl{4d})  increases in absolute value, whereas the \textsl{Y} atoms, 
 with the exception of Co, contribute small moments in both structure types. Even for the compounds that do not follow the 
SP rule (Mn$_2$NiGa, Mn$_2$CuGa, Mn$_2$RuGa, Mn$_2$RhGa, Mn$_2$PdGa), the trends 
 regarding the local moments are still valid. Regarding the large constant localized moment, 
 Mn is thus the only $3d$ transition metal element in inverse Heusler
 compounds that behaves like a rare-earth element.

The calculated exchange constants supply Curie temperatures in two different approximations that are likely to border the 
experimental values. Fig.~\ref{fig:Tc} can be separated into the L2$_1$ and X$_a$ parts. In L2$_1$, the Curie temperatures show 
only little variation. Starting with the materials with ${N_{\rm V}=24}$ (X$_a$ transition point),  care has to be taken with these values 
because of the occurrence of a phase change to tetragonal structures for some compounds. If the compounds remain in 
the cubic structure, the trend shown indicates an increase in the Curie temperature owing to the increase of the local moments and an increase 
in the exchange constants. The latter can be attributed to the increase in conduction electron concentration, which is the essence of the 
Zener--DeGennes\cite{Zener1951:ab,DeGennes1960:ab} exchange model for ferromagnetic metals. We believe the antiferromagnetic 
coupling of the nearest-neighbor Mn atoms is an atomic property of Mn in the elementary metal.

\acknowledgments
The authors gratefully acknowledge financial support from Project P~1.2-A of research unit FOR 1464~``ASPIMATT'') and the
European Research Council (ERC) ``Idea Heusler!''.

\bibliography{bibl}
\end{document}